
\normalbaselineskip=20pt
\baselineskip=20pt
\magnification=1200
\overfullrule=0pt
\hsize 16.0true cm
\vsize 22.0true cm
\nopagenumbers
\def\lsim{\mathrel{\rlap{\lower4pt\hbox{\hskip1pt$\sim$}}
    \raise1pt\hbox{$<$}}}         
\def\gsim{\mathrel{\rlap{\lower4pt\hbox{\hskip1pt$\sim$}}
    \raise1pt\hbox{$>$}}}         

\def\overleftrightarrow#1{\vbox{\ialign{##\crcr
    $\leftrightarrow$\crcr
    \noalign{\kern 1pt\nointerlineskip}
    $\hfil\displaystyle{#1}\hfil$\crcr}}}
\long\def\caption#1#2{{\setbox1=\hbox{#1\quad}\hbox{\copy1%
\vtop{\advance\hsize by -\wd1 \noindent #2}}}}

\centerline{\bf Symmetry Problems in}
\centerline{{\bf Low Energy Physics}\footnote*{Supported in part by the U.S.
Department of Energy}}
\vskip 18pt
\centerline{Ernest M. Henley}
\centerline{\it Department of Physics, Box 351560}
\centerline{\it University of Washington}
\centerline{\it Seattle, Washington  98195-1560  U.S.A.}
\vskip 24pt
\centerline{\bf ABSTRACT}

Some recent experimental and theoretical work on 1) charge symmetry-breaking,
2) parity non-conservation, and 3) searches for breaking of time
reversal invariance are reviewed.  The examples illustrate the uses of
symmetry to learn about underlying dynamics and/or structure.

\centerline{\bf INTRODUCTION}

Nuclei are known to be a superb laboratory for studies of symmetries and
symmetry-breaking as well as tests of our basic understanding of the
physical world.  It is easy to change mass, spin, isospin, charge, and
other properties of the nuclear target to allow us to probe various aspects
of basic theory.  Symmetries are particularly useful because they serve to
restrict the underlying dynamics, or, if the latter is known, allow a
determination of (unknown) structure.

In the time allotted to this talk, it clearly is not possible to discuss all
the symmetries useful in low energy physics.  I intend to concentrate on
charge symmetry, parity, and time reversal symmetries.  As we have heard at
this conference, there are many more symmetries one can discuss, such as chiral
symmetry and those that occur in the standard model.  Even within the above
restriction, I find it necessary to pick out a sample of the many interesting
features.

\centerline{\bf 1.  Charge Symmetry$^1$}

It is well known that hadronic (QCD) forces respect charge independence and
charge symmetry at low energies.  These symmetries are of particular interest
because their violation is small and can be studied both experimentally and
theoretically.  The violation occurs due to electromagnetic effects and the
mass difference of the up and down (d) quarks in the underlying QCD theory.
The violation of charge independence is of the order of a few percent.  It is
readily measurable in low energy nucleon-nucleon scattering, in the energy
spacings of isobaric analog states, and many other phenomena.$^1$

Recently, there has been more interest in the breaking of the looser charge
symmetry.  This symmetry does not require full rotational invariance in
isospin (charge) space, but only invariance under reflection in a plane
perpendicular to the charge (third component of isospin) axis.  The symmetry
violation is smaller than that of charge independence, but chiral
perturbation theory indicates that the violation is considerably enhanced
in processes that involve two neutral pions.$^2$  The nuclear forces which
break charge symmetry are of the form (classes III and IV),$^3$
$$V_{III}(1,2) = U_{III} [\tau _{3}(1)+\tau _{3}(2)]\eqno(1a)$$ and
$$\eqalignno{
V_{IV}(1,2)&= U_{IV}[\vec \sigma (1) - \vec \sigma (2)] \cdot \vec L
[\tau_3 (1) - \tau _3(2)]\cr
&+ U_{IV}^ \prime \vec \sigma (1) \times \vec \sigma (2) \cdot \vec L [\vec
\tau (1)
\times \vec \tau (2)] _3&(1b)}$$ where U is a space and spin dependent
interaction, $\vec L$ is the relative orbital angular momentum operator, $\vec
\sigma$ is a Pauli spin operator and $\vec \tau$ is the corresponding isospin
operator.  V$_{III}$ causes a difference between the nn and pp systems,
whereas V$_{IV}$ affects the n-p system.  One of the dominant contributors to
both V$_{III}$ and V$_{IV}$ is $\rho-\omega$ mixing.  It is thus not
surprising that there has been considerable recent interest in $\rho-\omega$
mixing.

This interest was sparked by Goldman, Henderson, and Thomas,$^4$ who
showed that a perturbative quark loop calculation predicted a large
momentum (q) dependence of the mixing matrix element.  The value of $\rho
-\omega$ mixing is known at the $\omega$-mass, where it has been measured in
the cross section for e$^+$ e$^-\rightarrow \pi^+\pi^-$; its value at
q$^2$ = m$^2_\omega$ is$^5 <\rho \mid$ H$_{CSB} \mid \omega >$
= -(4520 $\pm$ 600) MeV$^2$ and it arises mainly from the mass difference
m$_d-$ m$_u$ = $\Delta$m $\approx$ 4MeV.$^1$  Other methods have been used by a
number of authors to evaluate the off-the-mass shell dependence,$^6$ and all
find that for the momenta (q) relevant for nuclear forces (q$^2<$ 0) the mixing
matrix has changed sign (see Fig. 1).

\vskip 3.5truein
\noindent Fig. 1.  Momentum dependence of $\rho-\omega$ mixing.$^1$
\smallskip
\smallskip

\noindent This sign change gives rise to a much smaller $\rho-\omega$ mixing
potential and spoils the agreement of theory and experiment; examples are the
$^3$He-$^3$H mass difference due to V$_{III}$, the n-p polarization asymmetry
due to V$_{IV}$.$^{1,6}$  The sensitivity of this asymmetry to $\rho-\omega$
mixing is shown in Fig. 2.  Charge symmetry predicts equal analyzing powers
for the neutrons and protons,
$${\rm A}_n (\theta) ={\rm A}_p (\theta) \eqno(2)$$ in elastic scattering of
polarized particles.$^1$  Fig. 2 shows the sensitivity of the TRIUMF$^7$
(477 MeV) and IUCF (183 MeV)$^8$ experiments to $\rho-\omega$ mixing;$^1$ the
on-mass-shell value is used in this Figure.
\vskip 3truein
\noindent Fig. 2.  Sensitivity of asymmetry to $\rho-\omega$ mixing in
$\vec n - \vec p$ elastic scattering.
\smallskip
\noindent If the off-mass-shell extrapolation is used, then the agreement of
theory and experiment at 183 MeV would be absent, but the agreement at 477 MeV
would be maintained, as would that at 347 MeV, reported by Van Oers at this
conference.$^{11}$  What is missing?  For instance, it could be effects due to
the simultaneous exchange of a $\pi$ and $\gamma$.  However, Cohen and
Miller$^9$ and, more recently, Gardiner, Horowitz, and Pickarowicz$^{10}$ have
shown that the charge dependence of the $\rho$ and $\omega$ couplings to the
nucleons also need to be taken into account.  Indeed, the latter have used
vector dominance and $\Delta$m effects to show that this dependence gives rise
to a class IV force of the right magnitude and sign for the difference in n
and p analyzing powers in elastic $\vec n - \vec p$ scattering at 183 MeV.$^8$
It clearly would be of interest to resolve the uncertainty caused by the
various analyses.  Precise experiments of $\pi^+\pi^-$ production by pions on
protons at the $\omega$-mass may help to differentiate between the two
mechanisms.  If $\rho-\omega$ mixing occurs, then a $\pi^+\pi^-$ decay of the
$\omega$ becomes possible; see Fig. 3a.  It should be observed as a blip in
the $\pi^+\pi^-$-decay at the $\omega$-mass as seen in
$e^+ e^- \rightarrow \pi^+ \pi^-$.  On the other hand, a violation of isospin
at the $\omega$-p vertex will not lead to such a decay; see Fig. 3b.  Thus,
one can detect $\rho-\omega$ mixing at $q^2 = m_\omega^2$, but not isospin
nonconservation at the nucleon-$\omega^o$ vertex in this manner.  Both
mechanisms arise primarily from the up-down quark mass difference, $\delta$m.
For space-like values of q$^2$, no experiment can differentiate between these
two mechanisms because they are not really different.  This is illustrated in
Fig. 3c, where the $\rho-\omega$ mixing is regarded as a vertex correction for
$\omega$-exchange.  The reason this can be done is the finite size of the
hadrons, e.g., form factors at the nucleon-vector meson vertices.  To observe
isospin mixing at the vertex, I can only think of doing $\rho$ production from
a deuteron or $^4$He, d + d $\rightarrow  ^4$He  $\rho^o$ away from the
$\omega^o$ mass.  But this experiment will be difficult to analyze.  (See ref.
7 for a more extensive discussion.)
\vskip 3truein
\noindent Fig. 3.
(a) $\pi^- p \rightarrow \pi^+\pi^-$ at the $\omega$ mass;
(b) $\pi^- p \rightarrow n \omega^o$ with a charge symmetry-breaking vertex;
(c) $\rho^o-\omega^o$ mixing for q$^2 < 0$ regarded as a charge-symmetry
breaking $\omega$-N vertex.
\smallskip
\smallskip
\noindent How is the symmetry affected by the nuclear medium?  We know that
the nucleon mass is reduced, even at normal densities to about 3/4 of its
isolated value. Does this imply that the constituent quark mass is likewise
reduced?  Is $\rho-\omega$ mixing affected?  It is likely that many quantities
are affected by being placed in a nuclear medium.  The Nolen-Schiffer effect,
the increased mass difference of mirror nuclei over that expected from the n-p
mass difference and Coulomb effects has had a number of explanations.  Among
them are charge-symmetry-breaking effects, particularly $\rho-\omega$ mixing,
$^{12}$ and a decreased mass difference of the n-p in the nucleus.$^{13}$
Both of these effects depend on the effective up and down quark mass
difference.  Both mechanisms may be effective, and it will be interesting to
pursue other charge symmetry-breaking effects in nuclei.

In light nuclei, an example of a charge symmetry test is the forward-backward
asymmetry predicted by class IV forces in the reaction np
$\rightarrow$ d$\pi^o$.$^{14}$
$$\rm A_{fb}={{{d\sigma\over d\Omega}(\theta) -
{d\sigma\over d\Omega} (\pi-\theta)}
\over {{d\sigma\over d\Omega}(\theta) +
{{d\sigma}\over d\Omega}(\pi-\theta)}}\eqno(3)$$

The asymmetry is expected to be $\approx 10^{-3}-10^{-2}$ at $\approx$ 300
MeV, where a new experiment is planned at TRIUMF.  Another example is d+d
$\rightarrow$ $^4$He$+\pi^o$, which is forbidden by charge symmetry.  At
about 600 MeV, the magnitude expected, primarily from $\pi-\eta-\eta^\prime$
mixing, is $\approx$ 0.2 pb / sr.$^{15}$  This reaction may have been seen at
Saturne at 1.1 GeV with  $d\sigma / d\Omega = 0.97 \pm 0.20 \pm 0.15$ pb / sr
at $\theta_{cm}$ = 107$^\circ$.$^{16}$  However, both the experiment and
theory are difficult.  These are but two such tests of charge symmetry in
nuclei, which undoubtedly will be done more accurately in the future.

Charge symmetry-breaking can also be tested in hypernuclei$^1$ (see also A.
Gal at this conference).  So far, only the mass difference of the mirror pair
$^4$H$_\Lambda-^4$He$_\Lambda$ has been determined.$^{17}$ It would be
interesting to study the A-dependence to see whether there is a Nolen-Schiffer
type anomaly here and whether it can be explained due to $\Lambda^o-\Sigma^o$
mixing and/or other (i.e., mesonic) effects.

Charge symmetry does not apply in the weak interactions.  However, charge
symmetry can be and has been used to test features of the standard model and
other symmetries.  For instance, the conservation of the weak vector current
(CVC) has been tested in the A=12 nuclei to about 6\%$^{18}$ and in the A=8
multiplet (see Fig. 4) to about the same accuracy;  here one uses the symmetry
of the $^8$B and $^8$Li to make comparisons to the $^8$Be radiative decay.
Further, the measurements allow one to limit the magnitude of possible second
class currents which violate G-parity and/or time reversal invariance, and
could arise due to $\Delta$m.  No such currents were seen, with a lower limit
which is 1/2 of previous measurements (d$_{II}$ / Ac $\lsim$0.4).
\vskip 3truein
\noindent Fig. 4.  CVC test in A = 8 nuclei.
\smallskip
\smallskip

\centerline{\bf II.  Parity Nonconservation$^{19}$}

Parity nonconservation (PNC) can be used to test the standard model as
well as to obtain structural information.  I will give examples of both.

In discussions of PNC in nuclear forces, recent interest has centered
on two facets.  The first is to find neutral current effects at low
energies; the second is to understand experiments related to compound
nuclear formation.

PNC experiments in light nuclei are reasonably well understood.$^{19}$
However, neutral current effects have not yet been seen in non-leptonic
processes, despite considerable effort.  Let me remind you that nuclei allow
one to isolate neutral currents.  The parity-violating (pv) nuclear forces of
interest are isovector in nature.  These forces cannot come from the normal
charged currents, proportional to $\cos\theta_c$, where $\theta_c$ is the
Cabibbo angle, but only from strangeness-changing currents ($\Delta$S = +1 and
-1), proportional to $\sin\theta_c \approx$ 0.22 and from neutral currents.
$^{19}$

Furthermore, the isovector pv nuclear force is carried almost solely by the
pion; thus, it is a long-range force.  The pv pion-nucleon coupling due to
strangeness-changing (charged) currents is reduced by $\sin^2\theta_c$, and
thus is only $f_\pi^{(\pm)} \sim 4 \times 10^{-8}$.

\vskip 4truein
\noindent Fig. 5.  Parity tests in light nuclei.$^{31}$
\smallskip
\smallskip

By contrast, the best quark model calculation (DDH)$^{20}$ gives the coupling
due to neutral currents as $f_\pi^{(0)} \sim 5 \times 10^{-7}$ with large
errors.  (A more recent value by G.G. Feldman et al$^{21}$, based on the same
model, is $3 \times 10^{-7}$.)  Experiments in light nuclei allow one to
isolate the isovector pv-force in nuclei where levels of the same spin and
isospin differing by one unit are close together in energy.  Examples are
shown in Fig. 5.  Experiments in $^{18}$F show that $f_\pi^{(0)}$ is at least a
factor of three smaller than the ``best" DDH value.$^{20}$  Indeed, PNC due
to neutral currents has not been observed in nuclei (e.g., $^{18}$F).  As we
heard at this meeting,$^{22}$ QCD sum role calculations predict a $f_\pi^{(0)}$
that, due to a cancellation, is an order of magnitude maller than that given by
DDH.  This result agrees with an earlier chiral perturbative result of Kaiser
and Meissner.$^{23}$

If these calculations are correct, then the searches for neutral current
effects in low energy nuclear physics are extremely difficult, because the
resulting pv-force due to neutral currents is then of the same order of
magnitude as that due to the charged (strangeness-changing) current ones, and
the latter will mask the former.  It is only the difference of
$f_\pi^{(0)}+f_\pi^{(\pm)}$ from the readily calculable $f_\pi^{(\pm)}$, alone,
which will signal the presence of neutral currents.  It will be quite a while
until the accuracy of experiments and theory are up to this challenge.  This
situation will be made even worse if the nucleon contains strange quarks, as
suggested by a number of electron and muon deep inelastic scattering
experiments.  The effect of strange quarks has been examined by Kaplan and
Savage;$^{24}$ they could enhance the asymmetry due to the charged current pv
force.

In a continuing search for enhanced PNC effects, it was natural to turn to
compound nuclear resonances.$^{25}$  Epithermal polarized neutron scattering
at a compound nuclear {P}-wave resonance can give enhancements of many orders
of magnitude, to the pv asymmetry a
$$a ={{{\sigma_+ - \sigma_-} \over {\sigma_+ + \sigma_-}}}\eqno(4)$$
where $\sigma_+(\sigma_-)$ is the cross section for RH(LH) polarized neutrons.
The enhancement arises from a number of causes.  If we write
$$a \sim {{\sum_n {\langle f \mid T \mid S_n \rangle \langle  S_n \mid H_
{WK} \mid P \rangle}\over {E_P - E _S}}} \sim G_F m_\pi^2E \approx
10^{-7}E\eqno(5)$$
where P is a P wave and S its admixed opposite parity component, $a$ is
enhanced by E over its normal value $\sim G_F m_\pi^2$ by a number of factors.
 Here G$_F$ is the Fermi coupling constant and m$_\pi$ is the pion mass.
These factors are: (i) an enhanced transmission factor for S-waves over
P-waves, E$_1 \sim$ 1/kR$\sim 10^2-10^3$, where R is the nuclear radius and
k the wave number of the incident neutrons; (ii) a small energy denominator
$\Delta E \simeq E_P-E_S \approx$ 1 MeV / $\cal N$, where
${\cal N} \sim 10^4-10^6$ is the number of underlying levels of a resonance.
Although the matrix element $\langle\mid H_{Wk}\mid\rangle$ is suppressed by
$\cal N$, we obtain a net enhancement factor E$_2 \approx 10^2-10^3$ and thus
E = E$_1 E_2 = 10^4-10^6$.

This large enhancement means that asymmetries of several per cent can be
otained.  Such asymmetries have been seen in a number of experiments,$^{26}$
most recently at Los Alamos, e.g. on $^{232}$Th.  The initial results appeared
to indicate a non-statistical distribution of asymmetries, but these findings
have disappeared with more careful measurements.$^{26}$  An understanding of
the experimental results$^{25,26}$ is necessary in order to use the same
targets for time reversal tests.  Further developments will surely follow.

\vskip 3truein
\noindent Fig. 6.
(a) Standard PNC experiment in atoms;
(b) The smaller PNC test with a weak neutral axial nuclear current;
(c) The anapole moment contribution.
\smallskip
\smallskip

In semi-leptonic interactions, atomic PNC measurements and calculations have
now reached a level of accuracy of 1\%,$^{27}$ e.g. in $^{205}$Tl.  These
experiments depend on a weak axial electronic current and a weak vector
current coupling to the nucleus (Fig. 6a); the latter is enhanced by nuclear
coherence $\sim$ N, where N is the number of neutrons.  By contrast, the vector
electronic current and axial nuclear current (Fig. 6b) is not so enhanced,
because the axial current is proportional to the nuclear spin, which is due to
one, or at most a few, nucleons.  However, in heavy atoms this effect is
expected to be masked by the so-called nuclear anapole moment,$^{28}$ which
has yet to be seen.  The nuclear anapole moment is an axial coupling of a
photon to the nucleus (Fig. 6c), which arises from nuclear PNC.  It would be
interesting to determine whether the predicted masking actually occurs.

The experiments in Tl were just sufficiently accurate to have been able to see
the predicted anapole moment, but was not observed.  I am sure that further
improvements will allow one to test the theoretical calculations and it will be
interesting to see this novel effect.  It requires nuclei, because for a
nucleon or electron gauge invariance precludes its measurement.$^{29}$  The
effect there is of order G$_F \alpha$, where $\alpha$ is the fine structure
constant for gauge invariance.  Other diagrams of this order must be included.
However, in nuclei the anapole, e.g., due to positive pion exchange currents,
is enhanced by at least $\cal Z$, the nuclear charge.  This allows a
measurement of this strange coupling of the form$^{30,31}$
$$\langle p' \mid j _\mu^{em5}\mid p \rangle = \overline{u} (p \prime)
{(\gamma_\mu q^2 - 2Mq_\mu) {\gamma^5}\over M^2} u(p) a(q^2)\eqno(6)$$
where M is the nucleon mass,  q = p'- p, $u$ is a spinor, and
a(0) is the anapole moment.

At higher energies, PNC in electron scattering with polarized electrons can be
used to test the standard model {\it and} provide information on structure.
Let me give an example of the latter.$^{31}$  If strange quarks are present in
the nucleon, then there are two new and unmeasured form factors for the proton.
Of course, some strangeness is expected ($\sim$3\%)$^{32}$, but the SMC and
SLAC NMC measurements (as well as $\nu$ elastic scattering) indicated that
there could be as much as a 10-20\% polarized s$\overline{s}$ quarks presence
in the proton.$^{33}$  For elastic scattering of polarized electrons from
protons, the current can be written$^{31,34}$
$$J_\mu^{em} = \bar u \gamma_\mu u F_1^{em} (Q^2) +
\bar u \sigma _{\mu\nu} q^\nu u\kappa_pF_2^{em} (Q^2)$$

\vskip -35pt
$$\eqno(7)$$

\vskip -35pt
$$J_\mu^{wk} = \bar u \gamma_\mu u F_1^{wk} (Q^2) + \bar u \sigma_{\mu\nu}q^
\nu uF_2^{wk}(Q^2) + \bar u \gamma_\mu \gamma_5 uF_A^{wk} (Q^2) + \bar u
{q_\mu\over 2M} \gamma^5u F_P(Q^2)$$
If there are no strange quarks present, then CVC guarantees
$$F_1^{wk} = {1\over4}(1 - 4 \sin^2\theta_W) F_1^{em}$$

\vskip -35pt
$$\eqno(8)$$

\vskip -35pt
$$F_2^{wk} = [{1\over4}(1 - 4\sin^2\theta_W) - {\kappa _n\over{\kappa _p}}]
F_2^{em}$$
where $\kappa_n$ and $\kappa_p$ are the neutron and proton anomalous magnetic
moments, and $\theta$$_W$ is the Weinberg angle.  The observed asymmetry comes
from an interference of the electromagnetic and weak interactions of the
electron and proton.  On the other hand, if strange quarks are present, then
F$_1^{wk}(Q^2)$ may not = F$_1^{em}$(Q$^2$), and
$$F_2^{wk\prime} \approx F_2^{wk} + \kappa_SF_2^S\eqno(9)$$
where $\kappa_S$ is a new constant and F$_2^S$ a new form factor.

The reason for the new constant and form factor is that in the structure of the
proton, i.e. in hadronic interactions, the s quark is an isoscalar, but in the
weak interactions, it has weak isospin = 1/2; this mismatch is not present for
u and d quarks, which have the same weak and strong isospins.
For the axial current, we have
$$F_A^{wk\prime} = F_A^{wk} + {1\over4} g_A^S G^{wk}(Q^2)\eqno(10)$$
Whereas $F_A^{wk}(0) = {1\over4} g_A = -{1.26\over4}$,  $g_A^S$ is an
unconstrained and unknown constant and G$^{wk}$ an unmeasured form factor with
G$^{wk}$(0) = 1.  An experiment (SAMPLE) has been undertaken$^{35}$ at MIT to
determine $\kappa_S$ and quasi-elastic $\mu$ scattering on $^{12}$C has been
proposed$^{36}$ to determine g$_A^S$.  Further experiments are planned at
CEBAF,$^{37}$ and others have been proposed.$^{31,34}$  Again, nucleon and
nuclear targets are particularly helpful in determining some new nucleon
structure factors.

\centerline{\bf III.  Time Reversal Invariance$^{38}$ (TRI)}

Although CP violation was discovered over 30 years ago, its cause still has
not been established, despite considerable effort.  In nuclear physics, no
violation of TRI has been observed at a level of a few parts in 10$^4$.
In weak interactions, the $^8$Li and $^8$B beta decays to $^8$Be are being
used to improve these limits.  The term in the decay rate of
$\propto \vec\sigma \cdot \langle\vec J \rangle \times \hat{p}$
which is odd under both P and T transformations, has been sought.  Here
$\langle \vec J \rangle$ is the polarization of the nucleus; $\vec\sigma$ that
of the electron and $\hat{p}$ is a unit vector along a momentum [e.g., that of
one of the alpha particles from the $^8$Be decay or that of the electron
(positron)].  The coefficient R multiplying this term has been measured in
$^8$Li and reported at this conference to be $(4 \pm 35) \times 10^{-4}.
^{39}$  This limits the T-odd axial current matrix element $\langle p \mid j
\mu ^{(5)}\mid p \rangle = C_T \overline{u} (p\prime) i \sigma _{\mu\nu}
 \gamma_5 {q^\nu \over 2M}$ u (p) by $I_m C_T \lsim 0.010$.  A related
experiment to limit C$_T$ is being planned in Seattle.$^{40}$

\vskip 3truein
\noindent Fig. 7.  A time reversal invariance test with polarized neutrons
on a polarized target.
\smallskip
\smallskip

In addition, experiments can be carried out in heavy nuclei with polarized
epithermal neutrons.  In this case, the same compound nucleus enhancement
factors of $10^4-10^6$ which operate in the PNC experiments should be present.
A possible experiment$^{41}$ is to search for the (P-odd, T-odd) triple
correlation $\vec\sigma \cdot \langle \vec s \rangle \times \hat{p}$  or the
P-even, T-odd correlation $\vec \sigma \cdot \langle \vec s \rangle  \vec
\sigma \cdot \langle \vec s \rangle \times \hat{p}$, where $\hat{p}$ is a unit
vector along the incident momentum of the polarized neutrons scattered
coherently in the forward direction (see Fig. 7) from a polarized ($\langle
\vec J \rangle$) heavy nuclear target, such as $^{139}$La or $^{232}$Th.
Although the enhancement is likely to be insufficient to observe a TRI (and P)
violation, the effort needs to be and is being made.$^{41}$  Furthermore, the
accuracy on the limit of $\theta$ (see below) could rival that of searches for
a neutron dipole moment.

At this time, the most sensitive tests of TRI (and P) violation are searches
for electric dipole moments (d$_E$).  The electric dipole moment of the neutron
is now known to be$^{42} \lsim 8 \times 10^{-26}$ e-cm.  New efforts to
improve the limit are underway in Japan, Russia and at ILL, as reported at this
conference.$^{43}$  There also have been impressive improvements in searches
for d$_E$ of atoms; their accuracy now rivals or exceeds that of the neutron,
$^{44}$ with d$_E$ ($^{199}$Hg) 8$\times10^{-28}$e-cm.  In terms of the strong
CP parameter in the QCD Lagrangian, $\cal L$ (T-odd) = ${\alpha_S \over 8
\pi}\theta G_{\mu\nu} \tilde G^{u\nu}$,
where G is the gluonic field operator and $\tilde G$ its dual, these
experiments limit $\theta$ to $\theta \lsim 10^{-10}$.  Why $\theta$ should be
so small is not known.  Progress requires the observation of  a non-zero value
of a T-odd observable in a system other than $K^o - \overline K^o$.

\centerline{\bf CONCLUDING REMARKS}

Symmetries continue to be important in learning about the underlying dynamics
(i.e. interactions) and obtaining structural information.  In this talk I have
given several examples of the usefulness of studying small symmetry-breakings
that are of current and future interest.  There are many other ones at both
low and high energies.

\centerline{\bf REFERENCES}

\item{1.}  For a recent review, see G.A. Miller and W.T.H. Van Oers in
Symmetries and Fundamental Interactions Nuclei, ed. W. Haxton and E.M. Henley,
World Scientific (in press).

\item{2.}  U. Van Kolck (private communication).

\item{3.}  E.M. Henley and G.A. Miller, in Mesons in Nuclei, ed. M. Rho and
D.H. Wilkinson, North, Holland, Amsterdam (1979) 405.

\item{4.}  T. Goldman, J.A. Henderson, and A.W. Thomas, Few Body Systems
{\bf 12} (1992) 193.

\item{5.}  S.A. Coon and P.O. Barrett, Phys Rev {\bf C36} (1987) 489.

\item{6.}  J. Piekarewicz and A.G. Williams, Phys. Rev. {\bf C47} (1993) R2462;
G. Krein, A.W. Thomas, and A.G. Williams, Phys. Lett {\bf B317} (1993) 293; T.
Hatsuda et al, Phys. Rev. {\bf C49} (1994) 452; H.B. O'Connell et al, Phys.
Lett. (in press).

\item{7.}  T.D. Cohen and G.A. Miller, University of Washington preprint.

\item{8.}  S. Gardner, C.J. Horowitz, and J. Piekarewicz (Indiana Univ.
preprint IU/NTC 95-05).

\item{9.}  R. Abegg et al., Phys. Rev. Lett. {\bf 56} (1986) 2571; Phys. Rev.
{\bf D39} (1989) 2464.

\item{10.}  L.D. Knutson et al., Phys. Rev. Lett. {\bf 66} (1991) 1410.

\item{11.}  W.T.H. Van Oers (private communication and submitted for
publication).

\item{12.}  P.G. Blunden and M.J. Iqbal, Phys. Rev. Lett. {\bf B198} (1987) 14.

\item{13.}  E.M. Henley and G. Krein, Phys. Rev. Lett. {\bf 62} (1989) 2586.

\item{14.}  C.Y. Cheung, E.M. Henley and GA Miller, Phys. Rev. Lett. {\bf 43}
(1979) 1215; Nucl. Phys. {\bf A305} (1978) 342; {\bf A348} (1978) 365; J.A.
Niskanen, M. Sebestyen and A.W. Thomas, Phys. Rev. {\bf C38} (1988) 838.

\item{15.}  C.Y. Cheung, Phys. Rev. Lett. {\bf B119} (1984) 47.

\item{16.}  L. Goldzahl et al., Nucl. Phys. {\bf A533} (1991) 675.

\item{17.}  M. Juric et al., Nucl. Phys. {\bf B52} (1973) 1.

\item{18.}  L. De Braekeleer et al., Phys. Rev. {\bf C51} (1995) 2778.

\item{19.}  W. Haeberli and B.R. Holstein, in Symmetries and Fundamental
Interactions in Nuclei; loc. cit.; E.G. Adelberger and W.C. Haxton, Ann. Rev.
Nucl. Part. Sci. {\bf 35} (1985) 501.

\item{20.}  B. Desplanques, J.F. Donoghue, and B.R. Holstein, Ann. Phys. (New
York) {\bf 124} (1980) 449.

\item{21.}  G.B. Feldman et al., Phys. Rev. {\bf C43} (1991) 863.; S.A. Page
et al., Phys. Rev. {\bf C35} (1987) 1119; H.C. Evans et al, Phys. Rev. Lett.
{\bf 55} (1985) 791; M. Bini et al., Phys. Rev. Lett. {\bf 55} (1985) 795.

\item{22.}  W-Y. P. Hwang report at this conference; E.M. Henley and W-Y P.
Hwang and L.S. Kisslinger (preprint).

\item{23.}  N. Kaiser and U.G. Meissner, Nucl. Phys. {\bf A499} (1989) 699;
{\bf A510} (1990) 1648; U.G. Meissner, Mod. Phys. Lett. {\bf A5} (1990) 1703.

\item{24.}  D.M. Kaplan and M.J. Savage, Nucl. Phys. {\bf A556} (1993) 653.

\item{25.}  For a recent review, see V.V. Flambaum and G.F. Gribakin, Progr. in
Part. Nucl. Phys. {\bf 35} (1995) 423; M.B. Johnson and J.D. Bowman, Phys.
Rev. {\bf C51} (1995) 999.

\item{26.}  J.D. Bowman et al., Phys. Rev. Lett. {\bf 65}  (1990) 1992; C.M.
Grankle et al., Phys. Rev. Lett. {\bf 67} (1991) 564; Phys. Rev. {\bf C46}
(1992) 778.

\item{27.}  See e.g., P.A. Vetter et al, Phys. Rev. Lett. {\bf 74} (1995)
2658.

\item{28.}  Ya. B. Zeldovich, Zh. Eksp. Teor. Fiz {\bf 33} (1957) 1531
[transl. JETP {\bf 6} (1957) 943; E.M. Henley, W-Y P. Hwang and G.N. Epstein,
Phys. Rev. Lett. {\bf B36} (1978) 28; W.C. Haxton, E.M. Henley and M.. Musoff,
Phys. Rev. Lett. {\bf 63} (1989) 949; V.V. Flambaum, I.B. Khriplovich, and
O.P. Sushkov, Phys. Rev. Lett. {\bf 146B} (1984) 367.

\item{29.}  H. Czyz et al, Canad. J. Phys. {\bf 66} (1988) 132.

\item{30.}  W.C. Haxton, E.M. Henley, and M. J. Musolf, Phys. Rev. Lett.
{\bf 63} (1989) 949; V.V. Flambaum, I.B. Khriplovich, and O.P. Sushkov, Phys.
Rev. Lett. {\bf 146B} (1984) 367.

\item{31.}  For a recent review, see F. Boehm in Symmetries and Fundamental
Interactions in Nuclei, ed. W. Haxton and E.M. Henley, loc. cit.

\item{32.}  W. Koepf, E.M. Henley and S.J. Pollock, Phys. Rev. Lett. {\bf B288}
(1992) 11.

\item{33.}  D. Adams et al., (SMC), Phys. Rev. Lett. {\bf B329} (1994) 399;
K. Abe et al., (SLAC), Phys. Rev. Lett. {\bf 74} (1995) 346; L.A. Ahrens et
al., Phys. Rev. {\bf D35} (1987) 785.

\item{34.}  See e.g. E.M. Henley, Rev. Mex. Fis. {\bf 40} (1994) 31.

\item{35.}  MIT/Bates proposal 89-06 (1989); D.H. Beck, Phys. Rev. {\bf
D39} (1989) 3248; R.D. McKeown, Phys. Rev. Lett. {\bf B219} (1989); private
communcation.

\item{36.} G.T. Garvey at al., Phys. Rev. Lett. {\bf B289} (1992) 249; LSND
collaboration, LAMPF proposal 1173.

\item{37.}  D.H. Beck, private communication, and CEBAF proposal PR-91-017
{}.

\item{38.}  W. Koepf, E.M. Henley, and S.J. Pollock, Phys. Rev. Lett. {\bf
B288} (1992) 11.

\item{39.}  M. Allet et al., Phys. Rev. Lett. {\bf 68} (1992) 572; and J.
Sromicki at this conference.

\item{40.}  L. De Braeckeleer, private communication.

\item{41.}  L. Stodolsky, Phys. Rev. Lett. {\bf B172} (1986) 5; V. Hnizdo and
C.R. Gould, Phys. Rev. {\bf C49} (1994) 6K; J.J. Szymanski et al., Nucl. Phys.
{\bf A527} (1991) 701c; J.P. Sonderstrom et al., Phys. Rev. {\bf C38} (1988)
2424.

\item{42.}  K.F. Smith et al, Phys. Rev. Lett. {\bf B234} (1990) 191; I.S.
Alterev et al., Phys. Rev. Lett. {\bf B276} (1992) 242.

\item{43.}  A. Masaike at this conference and private communication.

\item{44.}  J.P. Jacobs et al., Phys. Rev. Lett. {\bf 71} (1993) 2383; and E.W.
Fortson (private communication).

\end{document}